# Parametric amplification of optical phonons

A. Cartella[1], T. F. Nova[1,2], M. Fechner[1], R. Merlin[3] and A. Cavalleri[1,2,4] *

[1]Max Planck Institute for the Structure and Dynamics of Matter, Hamburg, Germany

[2]The Hamburg Centre for Ultrafast Imaging, Hamburg, Germany

[3]Department of Physics, University of Michigan

[4]Department of Physics, Clarendon Laboratory, University of Oxford, Oxford, UK

*e-mail: andrea.cavalleri@mpsd.mpg.de

**Amplification of light through stimulated emission or nonlinear optical interactions has had a transformative impact on modern science and technology. The amplification of other bosonic excitations, like phonons in solids, is likely to open up new remarkable physical phenomena. Here, we report on an experimental demonstration of optical phonon amplification. A coherent mid-infrared optical field is used to drive large amplitude oscillations of the Si-C stretching mode in silicon carbide. Upon nonlinear phonon excitation, a second probe pulse experiences parametric optical gain at all wavelengths throughout the reststrahlen band, which reflects the amplification of optical-phonon fluctuations. Starting from first principle calculations, we show that the high-frequency dielectric permittivity and the phonon oscillator strength depend quadratically on the lattice coordinate. In the experimental conditions explored here, these oscillate then at twice the frequency of the optical field and provide a parametric drive for lattice fluctuations. Parametric gain in phononic four wave mixing is a generic mechanism that can be extended to all polar modes of solids, as a new means to control the kinetics of phase transitions, to amplify many body interactions or to control phonon-polariton waves.**

The amplification of acoustic[1,2] and optical[3] phonons under intense laser and magnetic fields has long been the subject of theoretical studies. However, a clear demonstration of phonon amplification has only been reported for acoustic modes in semiconductor superlattices driven by electrical currents[4]. On the other hand, the amplification of optical phonons, which are generally connected to crystal symmetries and to structural phase transitions, has not yet been demonstrated.

Here, we explore the nonlinear response of optically driven phonons in dielectrics, we discuss their amplification and their coupling to electromagnetic radiation.

Figure 1 depicts the nonlinear dependence of the polarization $P$ on the phonon displacement $Q$ and on an external electric field $E$ in a chain of Si and C atoms. The response of this idealized chain was computed by first principle calculations. The polar optical mode of this chain involves the relative displacement of the silicon (red) and carbon (blue) sublattices (Fig. 1a), reminiscent of the in-plane mode of hexagonal SiC (see below).

The first contribution to the nonlinear polarization, which we will refer to as $P_L$, bears on the effective dipolar charge $Z^*$. Such *Born effective charge*, defined as $Z^* = \frac{\partial P_L}{\partial Q}$, is approximated by a constant in the linear response regime ($P_L = Z^* Q$ ) but depends on $Q$ for large lattice distortions (see Figures 1b and 1c). For the chain of Fig. 1a and, generally, for most dielectrics, the Born effective charge depends quadratically on the lattice coordinate $Z^* = Z_0^* + \alpha Q^2$ (see Fig. 1c).

The second contribution to the nonlinear polarization emerges from the dielectric screening of the electric field $E$ by the electrons, giving the term $P_\infty = \varepsilon_0 \chi E = \epsilon_0 (\varepsilon_\infty - 1) E$.

Like the Born effective charge, the permittivity $\varepsilon_\infty$ is a constant for small lattice displacements, but becomes dependent on $Q$ when the lattice is strongly distorted. This effect is captured by the calculations of figure 1d for the Si-C chain, in which the slope $\chi = \varepsilon_\infty - 1$ of the polarization $P_\infty = \varepsilon_0 \chi E$ is shown to change at large values of $Q$. This second nonlinear term scales also quadratically with $Q$, as $\varepsilon_\infty = 1 + \chi = 1 + \frac{\partial P_\infty}{\partial E} = \varepsilon_{\infty,0} + \beta Q^2$ (see Fig. 1e).

Summarizing, the nonlinear polarization of a strongly driven optical mode in SiC (and a generic dielectric) includes two nonlinear corrections, both quadratic in Q, one to the effective dipolar charge $Z^*$ and one to the dielectric constant $\varepsilon_\infty$. Let us now consider the dynamical response of the lattice to an optical field $E = E_0 \sin(\omega t)$, tuned at or near the resonance associated with the transverse optical mode (see figure 2).

For small field amplitudes, $P = Z_0^* Q + \epsilon_0 (\varepsilon_{\infty,0} - 1) E$ and the time dependent phonon coordinate $Q(t)$ follows the familiar equation of motion of a periodically-driven damped oscillator $\ddot{Q} + \Gamma \dot{Q} + \Omega_{TO}^2 Q = Z^* E_0 \sin(\omega t)$, in which $\Gamma$ and $\Omega_{TO}$ denote damping and phonon frequency, respectively. In this case one obtains the familiar linear response expression $Q = Q_0 \sin(\omega t) \exp(-\Gamma t)$.

We first analyze the nonlinear response of $P_L$ alone, that is, when the Born effective charge (but not $\varepsilon_\infty$) depends on the lattice displacement ($\alpha \neq 0, \beta = 0$). The equation of motion is $\ddot{Q} + \Gamma \dot{Q} + \Omega_{TO}^2 Q = (Z_0^* + \alpha Q^2) E_0 \sin(\omega t)$. To leading order, the solution to this equation can be thought of as generating a set of harmonics in the force term. For oscillations in $Q$ at frequency $\omega$, the Born effective charge oscillates at $2\omega$, as $Z^* = Z_0^* + \alpha Q^2 \sim Z_0^* + \alpha Q_0^2 (\frac{1}{2} - $ $\frac{1}{2}\cos(2\omega t))$ (see Fig. 2). As a consequence, the driving force includes also a $3\omega$ component as $\alpha Q^2 E_0 \sin(\omega t) \sim \alpha Q_0^2(\frac{1}{2} - \frac{1}{2}\cos(2\omega t))E_0 \sin(\omega t)$.

More important is the nonlinear response of $P_\infty = \epsilon_0\left(\varepsilon_{\infty,0} + \beta Q^2 - 1\right)E$ to the driving field ($\alpha = 0, \beta \neq 0$). This term translates into a $+\beta Q^2 E$ correction to the polarization and, thus, to a change in the energy of the system $\Delta U = -PE = -\beta Q^2 E^2$. Hence, an additional force emerges on the oscillator $F_Q = -\frac{\partial U}{\partial Q} = +\beta E^2 Q$. The equation of motion can therefore be written as $\ddot{Q} + \Gamma\dot{Q} + \Omega(t)_{TO}^2 Q = Z^* E_0 \sin(\omega t)$, with a time dependent phonon eigenfrequency $\Omega_{TO} = \Omega_{TO,0} - \beta E(t)^2$. Because $\Omega_{TO}$ oscillates at frequency $2\omega$, the equation of motion is that of a forced parametric oscillator. Parametric amplification of lattice fluctuations $Q(t)$ are then expected due to the nonlinearity of $P_\infty$.

In this work, we experimentally validate the prediction of phonon amplification in bulk silicon carbide (polytype 4H, Fig. 3a). The eigenvector of the phonon studied here is shown in Fig. 3b, which displays motions of the Si and C atoms in opposite directions along one of the in-plane crystallographic axes. The equilibrium linear reflectivity is reproduced in Fig. 3c, displaying a 5 THz wide reststrahlen band, between $\Omega_{TO}$= 24 THz to $\Omega_{LO}$= 29 THz. Large amplitude oscillations of the lattice were driven with mid-infrared pulses, which were generated with two Optical Parametric Amplifiers (OPAs) and Difference Frequency Generation (DFG), powered by a Ti:Sa femtosecond laser at 1 KHz repetition rate. The pump pulses were tuned to $\Omega_{LO}$ = 29 THz. At these frequencies, the pump pulses were focused to obtain approximately 9 MV/cm field strengths. A second independently tunable pair of OPAs and DFG set-up was used to probe the spectral response of the same pumped resonance in reflection geometry. The time resolved reflectivity was recorded with broadband probe pulses centered at 26.5 THz, with spectral weight covering the whole reststrahlen band. As the probe pulses exhibited a stabilized carrier-envelope phase, the time dependent optical properties could be measured with sensitivity to both amplitude and phase by electro-optic sampling. To sample at this frequency, we compressed the pulses from the Titanium sapphire laser using a near-infrared non-collinear OPA, which generated pulses with approximately 20 fs pulse duration. The results of our pump probe experiments are reported in figure 4, in which we plot the wavelength dependent reflectivity after excitation with pump pulses.

For pump electric fields $E_0 > 4$ MV/cm, the reflectivity in the reststrahlen band was observed to become larger than $R=1$, reaching the values of $R{\sim}1.15$ and evidencing amplification. This feature, emphasized in red in Fig. 4(a-d), developed at the earliest time delays and at the center of the reststrahlen band, broadening in frequency and persisting for longer time delays as the pump field was increased towards 9 MV/cm. As depicted in Fig. 4e, the reflectivity increases throughout the reststrahlen band, scaling quadratically with the pump electric field, as shown in Fig. 4f. These observations suggest that for large coherent excitations of the phonon, the probe electric field is amplified and, by extension, the lattice coordinate $Q$.

Numerical simulations of the optical response under the conditions of the experiment were used to analyze the results above and to validate amplification of both $E$ and $Q$. Starting from Maxwell's equations, we considered the interaction of electromagnetic transients of arbitrary shape and amplitude with the SiC crystal. In these calculations, which are discussed in the Methods section, we considered both pump (strong) and probe (weak) pulses and included the quadratic dependence of both the Born effective charge $Z^*$ and dielectric constant $\varepsilon_\infty$ on the phonon coordinate $Q$, as obtained from Density Functional Theory (DFT) analysis of the Si-C lattice.

The calculated $\alpha$ and $\beta$ coefficients were adjusted to fit the experimental data. These simulations reproduced well the main features of the time-delay-dependent and frequency-dependent reflectivity response measured experimentally, as reported in Fig. 5(a-d). For pump fields in excess of 4 MV/cm, the simulations predict $R$>1. Precisely as observed in the experiments, the frequency-dependent profile of the amplification ($R$>1) emerges from the center of the reststrahlen band and expands both in frequency and time as the pump field increases (Fig. 5e). The calculated maximum reflectivity also scales quadratically with the pump peak electric field (Fig. 5f), in agreement with the experiments. The simulations further confirm that not only the probe electric field (Fig. 6a), but also the oscillations in the phonon coordinate $Q$(Fig. 6b) are amplified. In particular, Fig. 6b displays the time dependent $Q(t)$ with (red) and without (black) excitation, highlighting phonon amplification.

The idea that lattice fluctuations can be amplified parametrically through the nonlinear response of the lattice is relevant in more than one area. For example, in the context of light enhanced superconductivity in cuprates[5–7] and in the doped fullerites[8], a recent theory has suggested that parametric amplification of pairs of squeezed phonons may enhance the superconducting instability[9,10]. Integral to these conjectures is the ability to parametrically amplify phonons by four wave vibrational mixing. The mechanism discussed here naturally extends to these conditions.

The present results are also connected to previous studies in charge density wave systems, in which parametric amplification of the order-parameter phase mode, driven by large amplitude coherent excitations of the amplitude mode has been discussed[11].

Finally, the physics of phonon amplification discussed here could be immediately extended to the manipulation of phonon-polaritons waves, which are of interest to information transport on sub-wavelength lengthscales[12]. The ability to control the properties and amplitude of phonon-polaritons may for example lead to tunable meta-lenses for the phonon field, or to many other extensions of optoelectronic manipulation to "polaritonics".

# Methods

## Experimental set-up

Laser pulses with 5.5 mJ energy, 800 nm wavelength and 100 fs duration at 1 KHz repetition rate from a commercial Ti:Sapphire regenerative amplifier were split into three parts (80%, 15%, 5%).

The 80% beam was used to pump a pair of two-stage optical parametric amplifiers (OPAs), seeded by the same white light continuum (WLC) and delivering phase-locked signal output pulses in the near infrared (300 µJ, 70 fs duration, tuned at 1.46 and 1.28 µm wavelength, respectively). The difference frequency generation (DFG) in a GaSe crystal between these signals delivered carrier envelope phase (CEP) stable[13] mid-infrared pump pulses of 130 fs duration, 29 THz center frequency, and up to 10 µJ energy. These pulses were transmitted through a pair of KRS-5 broadband wire grid polarizers (used to vary their intensity) and focused at normal incidence onto the sample with a 150mm effective focal length off-axis parabolic mirror. The beam diameter at the sample position, measured by its transmission through a calibrated pin-hole, was about 240 µm, yielding to a maximum fluence of 13 mJ/cm$^2$, corresponding to 8.68 MV/cm peak electric field.

The 15% beam was used to pump a similar setup, with two OPAs seeded by the same WLC and tuned at 1.45 µm and 1.29 µm, respectively. The DFG between these signals produced mid-infrared probe transients at 26.5 THz, with 100 fs duration and energy of 0.6 µJ. These were attenuated with a pair of gold wire grid polarizers and focused on the sample at 15 degrees from normal incidence. The probe spot size on the sample, also measured by its transmission via a calibrated pin-hole, was 170 μm. The reflected pulses were then collimated and steered towards the electro-optic sampling (EOS) setup[14], where their electric field profile could be measured.

The 5% beam was frequency doubled and used to pump a non-collinear optical parametric amplifier (NOPA)[15], delivering near-infrared pulses with a spectrum centered at 900nm, and with a 60nm bandwidth. These pulses were then compressed down to approximately 20 fs duration by several bounces on a chirp-mirrors pair, and used as a gate in the EOS setup. A schematic representation of the experimental set-up can be found in Supplementary Information 1.

To accurately measure the optical properties of the sample, a double optical chopping scheme was used[16]. This allowed for the simultaneous measurement of the equilibrium and pump-induced reflected electric fields, and to calculate $\Delta E(t,\tau)/E(t,\tau)$, where $t$ is the time delay between the pump and the probe pulses and $\tau$ the EOS time coordinate. From this ratio, the complex reflectance $\Delta r(t,\omega)/r(t,\omega)$ was calculated and added to the equilibrium one (retrieved from literature data[17]), thereby providing the complete optical properties of the sample as a function of both pump-probe delay and frequency. For more details on this topic, we refer the readers to Supplementary Information 2.

All the experiments were performed at room temperature.

**First principle calculations of Z\* and $\varepsilon_\infty$ as a function of displacement**

To explore the changes of the mode effective charge and $\varepsilon_\infty$ we performed first-principle computations within the framework of density functional theory (DFT). All our computations were carried out using DFT within a pseudopotential scheme as implemented in the Quantum Espresso code[18]. We used pseudopotentials generated by the projected augmented wave (PAW)[19] scheme for Si and C which contain as valence states the $3p^2\,3s^2$ and $2p^2\,2s^2$ electrons for each element, respectively. As numerical parameters, we applied a cutoff energy for the plane wave expansion of 75 Rydberg and 400 Rydberg for the charge density. For all computations, we sampled the Brillouin zone by a 21x21x7 k-point mesh generated with the Monkhorst and Pack scheme[20] and reiterated total energy calculations until the total energy changed less than $10^{-10}$ Rydberg. Before calculating phonon-modes and Born-effective charges of SiC, we structurally relaxed the unit-cell for forces and pressure below a threshold of 5 $\mu Ry/a_0$. We then performed density functional perturbation theory (DFPT)[21] calculations to obtain the phonon modes eigenvectors and frequencies. Finally, we utilized frozen phonon calculations, incorporating applied electric fields in the framework of the modern theory of polarization[22] to compute the Born-effective charges, the electronic susceptibility and their modulation upon structural deformation. For further details on this topic we refer the reader to Supplementary Information 3.

**Dynamical response simulations**

The time-dependent and frequency-dependent reflectivity of Fig. 5, as well as the $E$ and $Q$ amplification of Fig. 6, were calculated solving Maxwell's equations in space and time (discretized with the Yee grid[23]) with the one-dimensional finite difference time domain method (1D-FDTD)[24]. The relevant IR-active phonon resonance of SiC was implemented with an harmonic oscillator whose parameters (DC and high frequency dielectric permittivity $\varepsilon_0$ and $\varepsilon_\infty$, eigenfrequency $\Omega_{TO}$ and damping $\Gamma$) were extracted from a Lorentz fit to the equilibrium reflectivity data[17]. The $Q$-dependence of $Z^*$ and $\varepsilon_\infty$ obtained from first-principles computations (see previous methods section), were implemented in the FDTD loop with additional $\alpha$ and $\beta$ terms in both the polarization and oscillator equations as shown in the main text. These simulations allowed us to calculate the reflected probe electric field, both at equilibrium and after the sample was excited by an intense pump pulse. From these fields we calculated $\Delta E(t,\tau)/E(t,\tau)$, therefore directly replicating the experiments. The simulated electric field transients were then analyzed in the same way as the experimental ones, to directly compare the results. More information on this topic can be found in Supplementary Information 4.

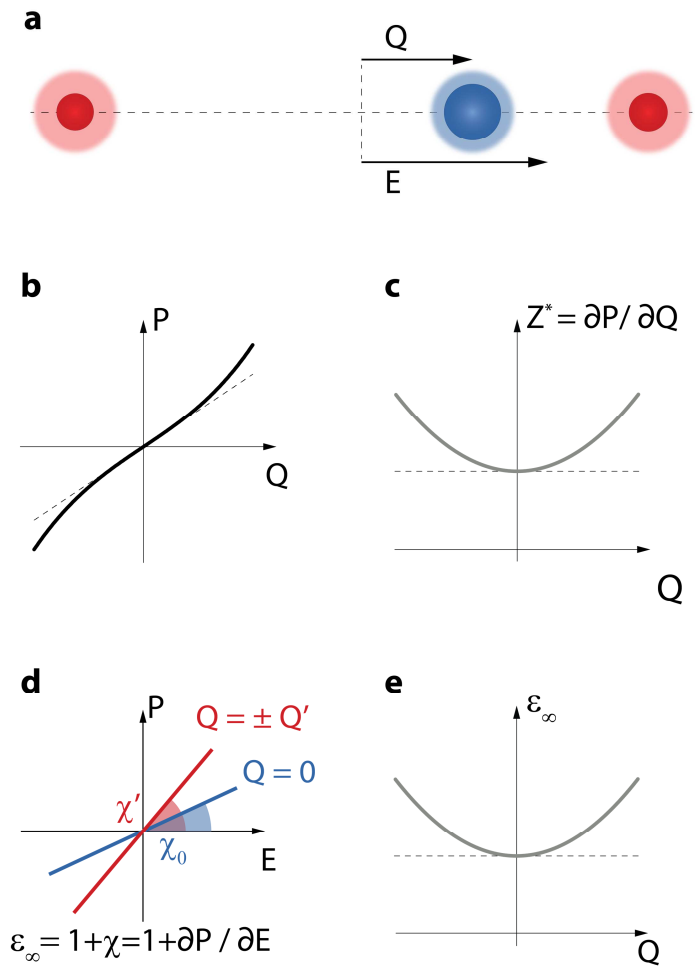

**Figure 1 | Polarization in a diatomic chain of Si and C atoms as a function of Si displacement and electric field**. **a**, Chain of Silicon (blue) and Carbon (red) atoms. At equilibrium, the positive ions are surrounded by negative electronic clouds, arranged in space so that no polarization is present. If an ion is displaced, a polarization $P$ is created along the chain. The system can be described by effective positive and negative ionic charges $Z^*$, depicted as shaded areas around the ions. **b,c**, Lattice polarization along the chain (c) and effective charge $Z^*$ (d) as a function of $Q$, resulting from first principle DFT calculations. For large displacements, the polarization is not linear in $Q$ anymore, and the effective charge is increased. The first expansion of $Z^*$ in $Q$ is parabolic (grey line). **d,e**, Non-resonant contribution to the polarization as a function of E for different values of Q (d) and dielectric constant $\varepsilon_\infty$ as a function of Q (e) resulting from first principle DFT calculations.

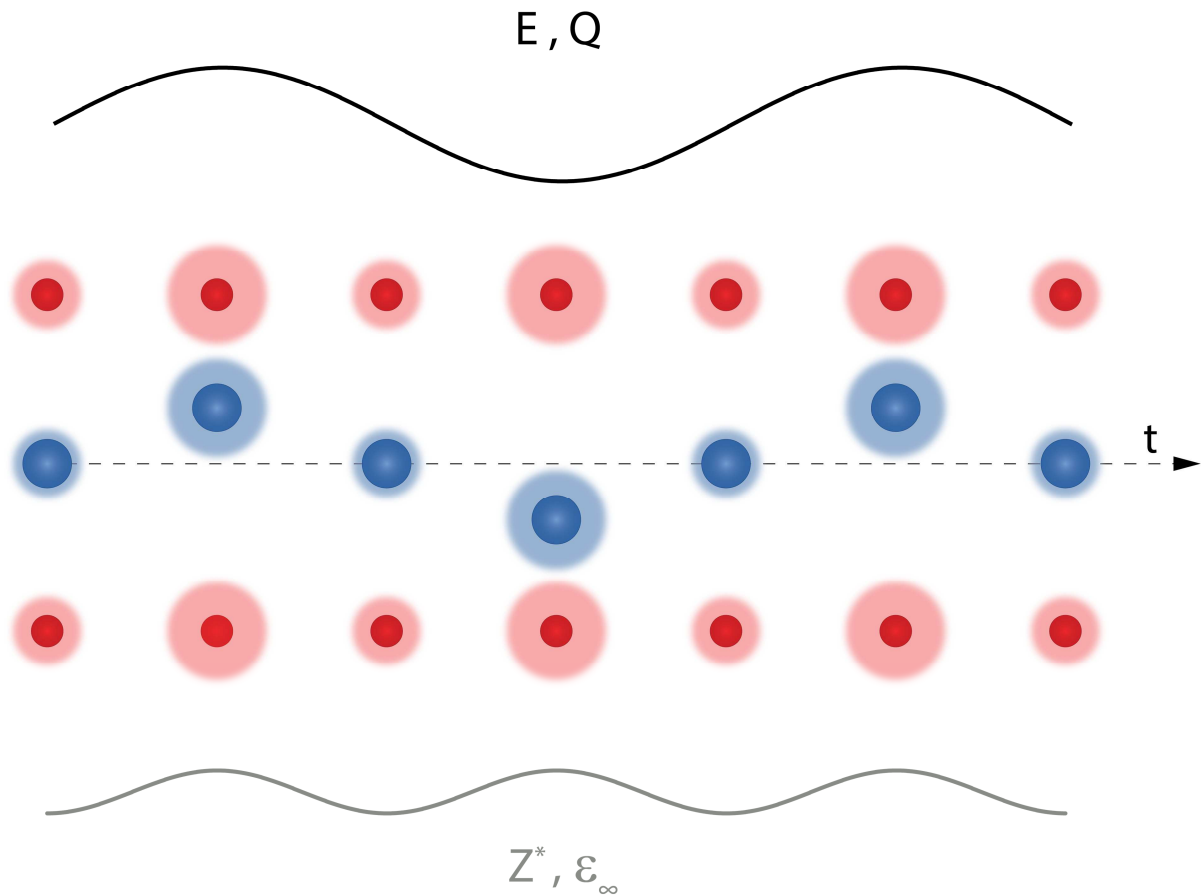

**Figure 2 | Dynamical response of the diatomic chain to a large amplitude periodic driving of the phonon.** For an applied electric field $E = E_o \sin \omega t$ (black line), the phonon coordinate Q oscillates as $Q = Q_o \sin \omega t$. Due to their quadratic dependence on Q, the effective charge $Z^*$ and the dielectric constant $\varepsilon_\infty$ (grey line) are oscillating at frequency $2\omega$.

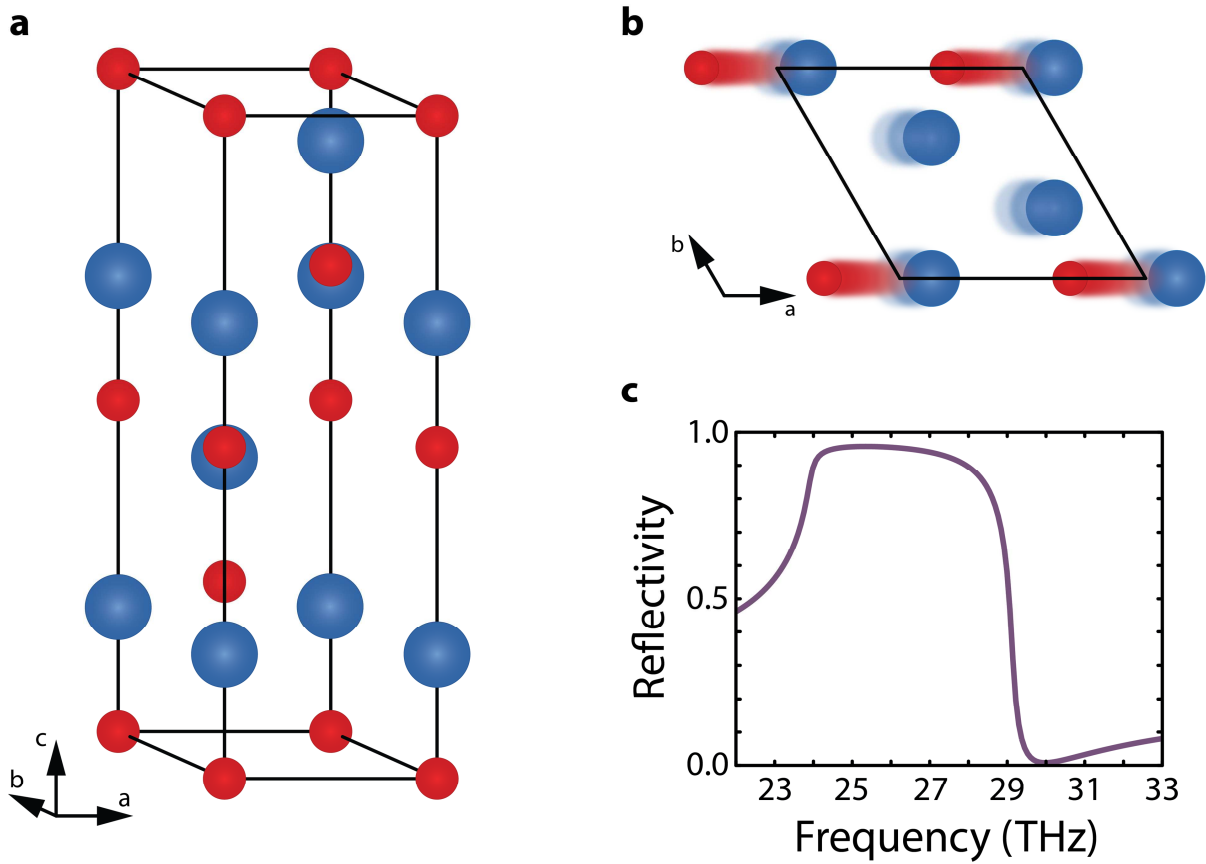

**Figure 3 | Silicon carbide structure and optical properties. a**, Crystal structure of SiC, polytipe 4H (space group $C^4_{6v}$-P6$_3$mc). Si atoms in blue, C atoms in red. **b**, Eigenvectos of the infrared actived mode excited by the pump pulse (E$_u$ symmetry). **c**, Reflectivity at equilibrium associated to the driven mode (data from literature[17]).

# Experiment

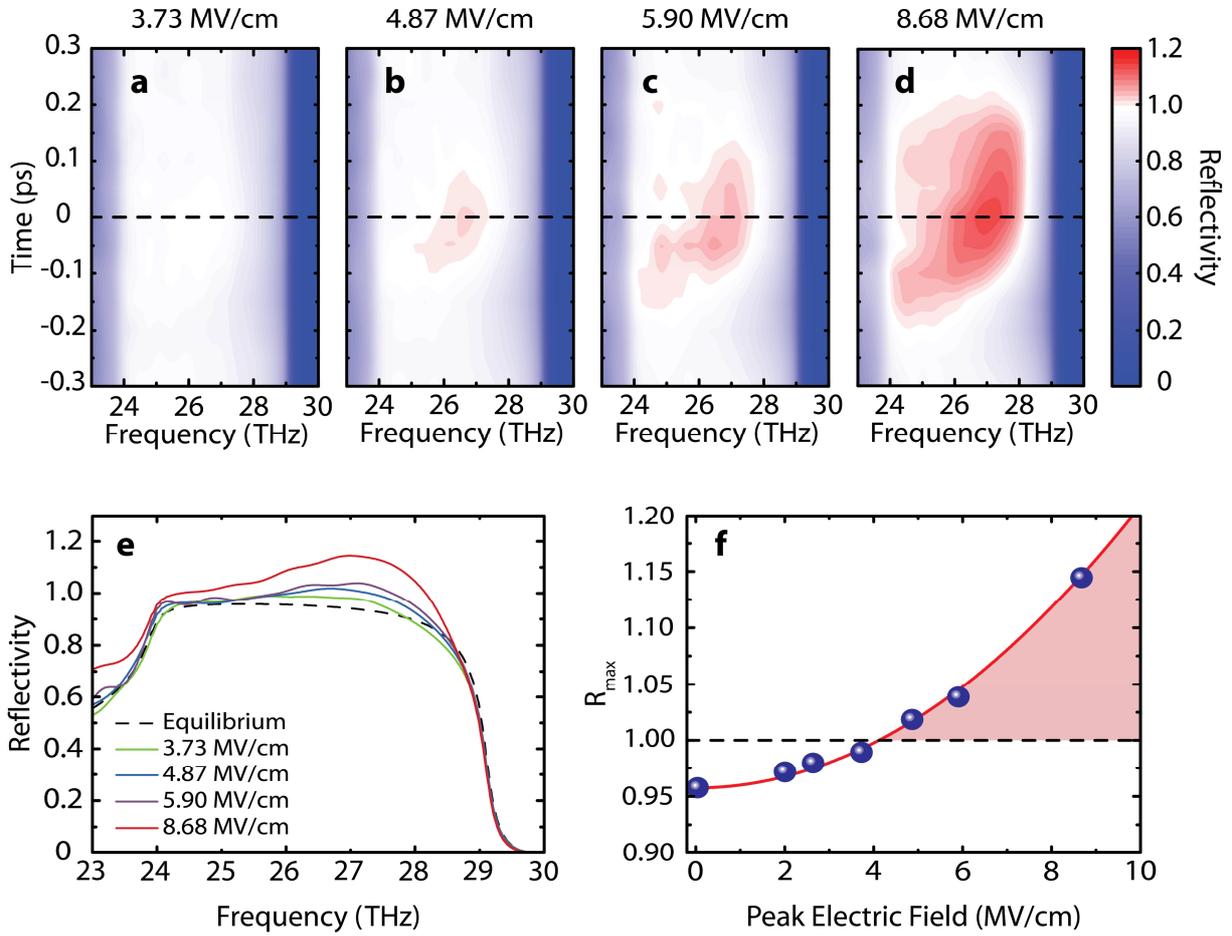

**Figure 4 | Experimental results. a-d**, Time-delay-dependent and frequency-dependent measured reflectivity $R(t, \omega)$ for driving peak electric fields of (a) 3.73, (b) 4.87, (c) 5.90 and (d) 8.68 MV/cm. The color scale is chosen to emphasize in red the regions where the reflectivity is greater than one. The horizontal dashed lines indicate the time delay at which the line-cuts are displayed in panel e. This is the delay at which the highest reflectivity was measured for the highest peak electric field (panel d). **e**, Frequency-dependent measured reflectivity at the maximum of the pump-probe response for different driving peak electric fields, compared to the equilibrium one (dashed line). **f**, Peak field dependnece of the maximum measured reflectivity. In this plot, two additional data points are shown, corresponding to 2.01 and 2.64 MV/cm. The red line depicts a parabolic fit to the data.

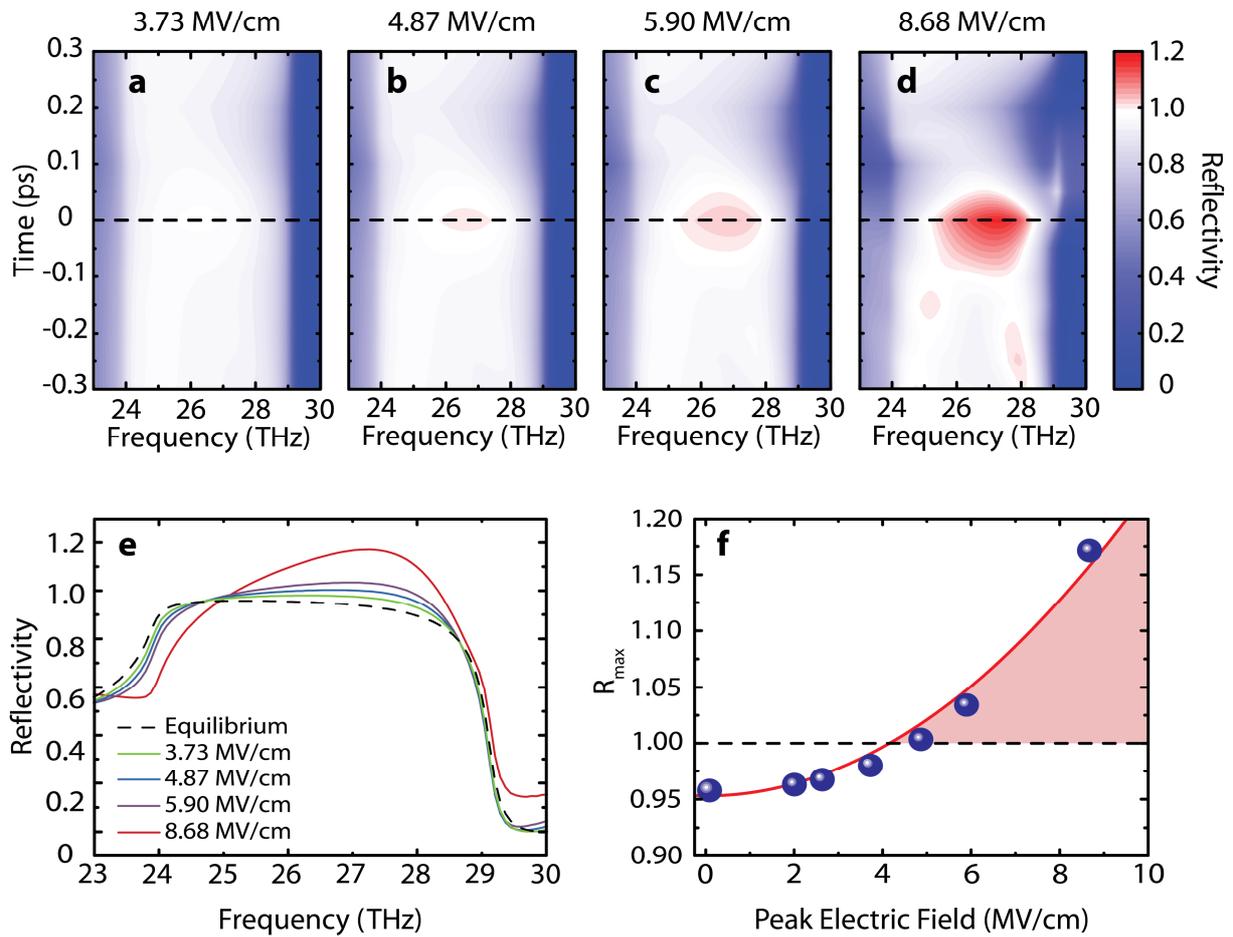

**Figure 5 | Simulations results. a-d**, Time-delay-dependent and frequency-dependent simulated reflectivity $R(t, \omega)$ for driving peak electric fields of (a) 3.73, (b) 4.87, (c) 5.90 and (d) 8.68 MV/cm. The color scale is chosen to emphasize in red the regions where the reflectivity is greater than one. The horizontal dashed lines indicate the time delay at which the line-cuts are displayed in panel e. **e**, Frequency-dependent calculated reflectivity at the maximum of the pump-probe response for different driving peak electric fields, compared to the equilibrium one (dashed line). **f**, Peak field dependnece of the maximum calculated reflectivity. In this plot, two additional data points are shown, corresponding to 2.01 and 2.64 MV/cm. The red line depicts a parabolic fit to the data.

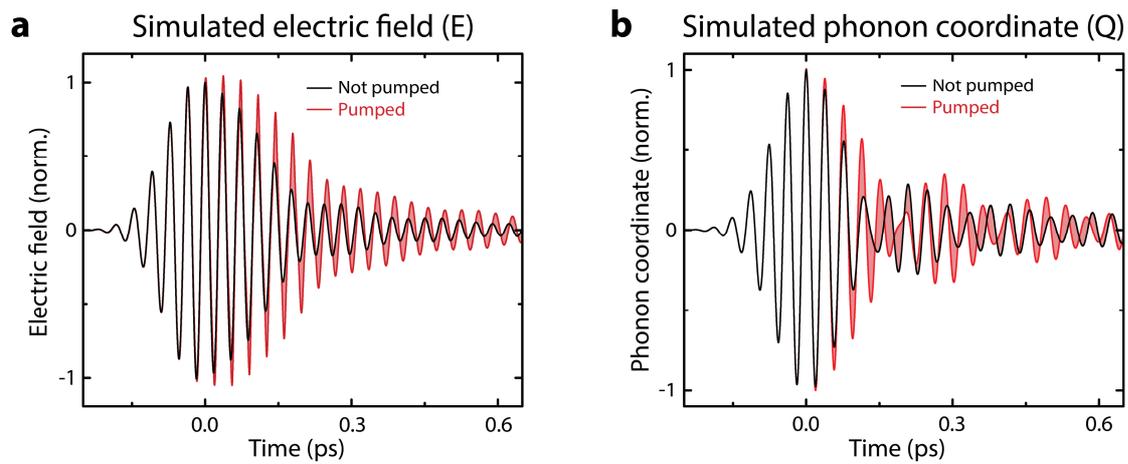

**Figure 6 | Probe pulse electric field and phonon amplification.** Simulatd electric field $E$ (panel **a**) and oscillations of the phonon coordinate $Q$ (panel **b**) at the sample surface, driven by a weak probe pulse, with (red solid lines) and without (black solid lines) pump. The shaded areas highlight the amplification.

# Parametric amplification of optical phonons

## *Supplementary Material*

A. Cartella[1], T. F. Nova[1,2], M. Fechner[1], R. Merlin[3] and A. Cavalleri[1,2,4] *

[1]Max Planck Institute for the Structure and Dynamics of Matter, Hamburg, Germany

[2]The Hamburg Centre for Ultrafast Imaging, Hamburg, Germany

[3]Department of Physics, University of Michigan

[4]Department of Physics, Clarendon Laboratory, University of Oxford, Oxford, UK

*e-mail: andrea.cavalleri@mpsd.mpg.de

## S1 – Experimental set-up

Here we report a schematic representation of the experimental set-up described in the Methods section.

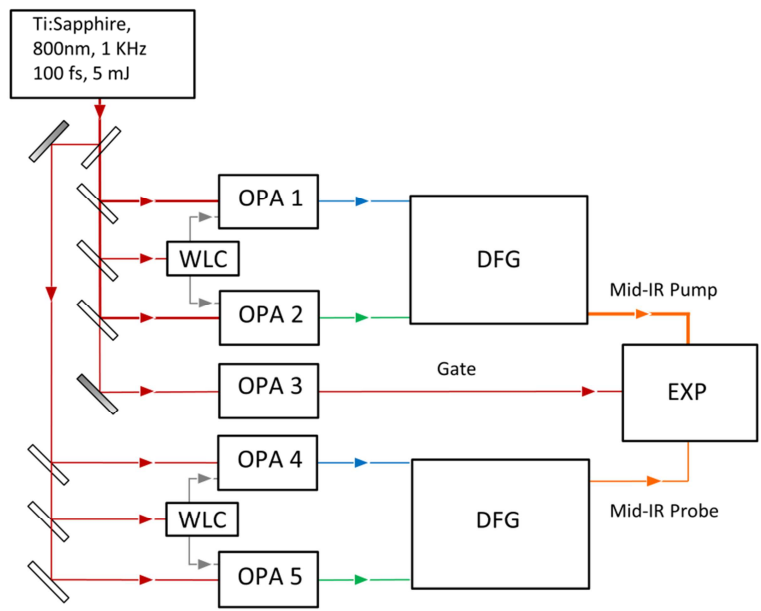

**Figure S1-1 | Schematic representation of the set-up used to generate the pump, probe and gate pulses used in the experiment**. Optical parametric amplifiers (OPA), white light continuum generation (WLC), difference frequency generation (DFG), experiment (EXP).

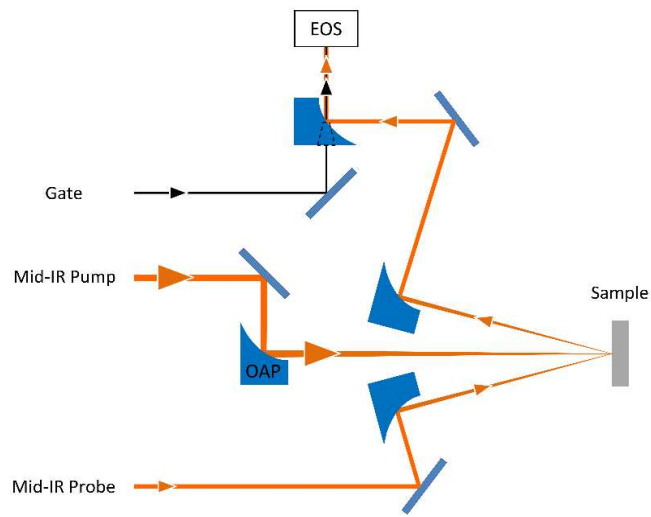

**Figure S1-2 |** Schematic representation of the experiment geometry. Off-axis parabolic mirror (OAP), Electro Optic Sampling (EOS).

## S2 – Data Analysis

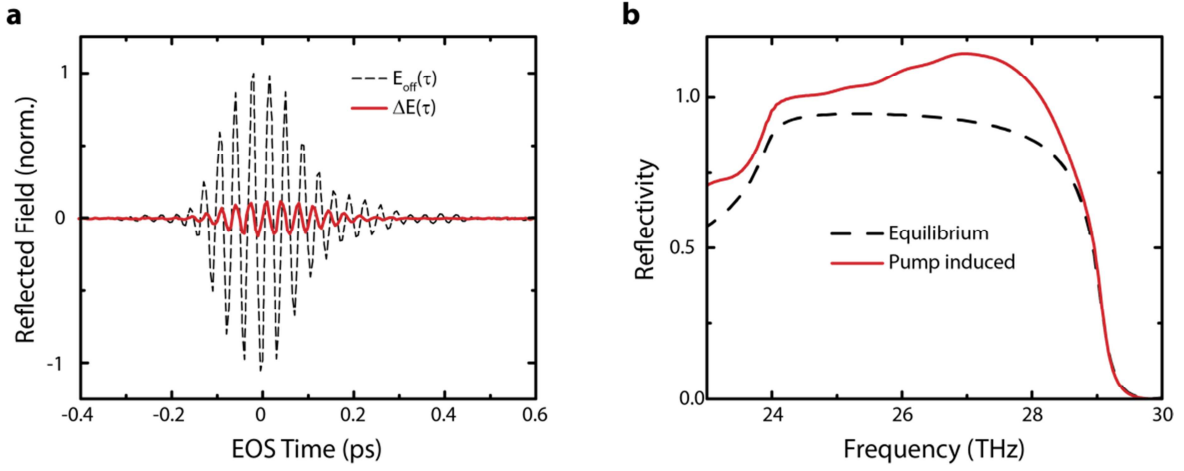

**Figure S2 | Raw data and analysis. a,** Raw data. The dashed line represents the probe electric field $E_{off}(\tau)$ reflected from the sample at equilibrium. The red line represents the simultaneously measured change $\Delta E(\tau)$ induced by the pump pulse. **b,** Equilibrium (dashed line) and pump-induced reflectivity (red line) calculated from the data in panel a.

In Fig. S2a we show the raw data for the measured probe pulses, after reflection from the sample, at a fixed pump-probe delay t and as a function of the EOS time coordinate $\tau$. For each pump-probe delay $t$, both the equilibrium reflected pulse $E_{off}(\tau)$ and the pump-induced change $\Delta E(\tau)$ were measured simultaneously. This was achieved by chopping both the pump and the probe beams, and measuring the reflected electric field with two lock-in amplifiers connected in parallel. By scanning the pump-probe delay, both $E_{off}(t,\tau)$ and $\Delta E(t,\tau)$ could be measured simultaneously. The frequency resolved data $E_{off}(t,\omega)$ and $\Delta E(t,\omega)$ could be reconstructed by Fourier transforming the measured traces along the EOS coordinate $\tau$. The reflectivity could then be calculated recalling that $\frac{\Delta E(t,\omega)}{E_{off}(t,\omega)} = \frac{\Delta r(t,\omega)}{r_{off}(t,\omega)}$, where $r(t,\omega)$ was the reflection coefficient. Because the equilibrium property $r_{off}(\omega)$ is known, the pump-induced reflection coefficient $r_{on}(\omega) = r_{off}(\omega)(1 + \frac{\Delta r(t,\omega)}{r_{off}(t,\omega)})$ could be calculated, as well as the reflectivity $R_{on}(\omega)$=$|r_{on}(\omega)|^2$. Figure S2b shows the reflectivity calculated from the data of Fig. S2a.

**S3 – DFT calculations**

We used first-principle calculations to reconstruct the change in dielectric properties of SiC due to direct phonon excitation. The numerical details of our computations are listed in the methods section. The basis of our study is the Hamiltonian:

$$H = \frac{\Omega^2}{2}Q^2 - \frac{\epsilon_0(\epsilon(\infty) - 1)}{2}E^2 - \frac{Z^*}{\sqrt{\mu\, V}}QE \quad , \qquad \text{(S3.1)}$$

where $\Omega$ is the polar-phonon frequency, $Z^*$ is the associated effective charge and $\epsilon(\infty)$ is the dielectric response at high frequencies. $Q$ and $E$ are the phonon mode coordinate and electric field amplitude, respectively, and $\epsilon_0$ is the vacuum permittivity. $V$ is the unit cell volume and $\mu$ is the mode's effective mass. For small phonon amplitudes the equilibrium optical properties are well described by this Hamiltonian. However, for large driving fields it has to be modified to take into account higher order effects. We therefore expand equation S3.1 up to 4th order considering all the symmetry allowed terms in $Q$ and $E$. The new Hamiltonian reads:

$$H = \left(\frac{\Omega^2}{2} + \phi Q^2\right)Q^2 - \left(\frac{\epsilon_0(\epsilon(\infty) - 1)}{2} + \beta Q^2 + \xi E^2\right)E^2$$
$$- \left(\frac{Z^*}{\sqrt{\mu\, V}} + \alpha Q^2 + \theta E^2\right)QE \quad . \qquad \text{(S3.2)}$$

Where $\phi, \beta, \xi, \alpha$ and $\theta$ are dimensional constants that describe the strength of the nonlinear corrections. Please note that equation S3.2 has been written to highlight in parenthesis the corrections to the terms of equation S3.1.

To estimate the relative strength of the expansion coefficients we performed total energy density functional theory calculations, which are mapped on Eqn. (S3.2).

The starting points of the computations are the ground-state dielectric properties of 4H-SiC. First, we did a structural minimization of the 4H-SiC unit cell to obtain a force/stress free DFT reference structure. During the minimization, we constrained the 4H-SiC to the experimentally determined[1] P63mc symmetry. The obtained structure parameters are listed in Tab. S3.1.

| Lattice constant | | DFT | Expt. |
|---|---|---|---|
| a (Å) | | 3.06 | 3.08 |
| c/a | | 3.272 | 3.26 |
| Atom | Wyckoff position | DFT | Expt. |
| Si | 2a | 0.187 | 0.187 |
| Si | 2b | 0.437 | 0.437 |
| C | 2a | 0.000 | 0.000 |
| C | 2b | 0.250 | 0.250 |

Table S3.1: Experimental[1] and calculated (DFT) lattice constants and atomic positions for 4H-SiC. The Wyckoff positions are given according to the P63mc space group.

Starting from this structure, we performed density functional perturbation theory calculations[2] to determine the phonon frequencies, $\epsilon(\infty)$ and the Born effective charges. The calculated frequency of the excited E2 phonon mode is 23.3 THz. The eigenvector of this mode displaces Si and C in opposite directions along the hexagonal *a*-axis. The calculated displacement is 6 *pm* and 12 *pm* for Si and C, respectively.

To determine the coefficients in Eqn. (S3.2) we then performed frozen phonon calculations including applied electric fields in the framework of the modern theory of electric polarization. We thereby displaced the structure along the E2 phonon eigenvector and computed the total energy for different applied electric fields. The phonon amplitudes ranged from -1 to 1 times the eigenvector and the electric field from -10 to 10 MV/cm. Finally, we did least mean square minimization of the resulting 2-dimensional energy landscape. Please note that the only parameters allowed to vary within the fit were the additional appearing coefficients in Eqn. (S3.2). Tab. S3.2 summarizes the calculated coefficients, including the equilibrium properties.

To identify the most relevant changes in the energy landscape due to structural modulation we computed the total energy for an electric field of 8 MV/cm and the corresponding phonon amplitude. We then calculated the energy contribution of each term in percent, clearly showing the leading contribution of the $\alpha$ and $\beta$ terms among the expansion ones.

| constant | coefficient | value | Relative energy contribution [%] |
|---|---|---|---|
| $\Omega$ | $Q^2$ | 23 THz | 92.6 |
| $\epsilon_\infty$ | $-E^2$ | 5.91 | 0.2 |
| $Z^*$ | $-QE$ | 2.61 e | 6.7 |
| $\alpha$ | $-Q^3E$ | $3.0 \; 10^5 \; eV/(u^{\frac{3}{2}}\sqrt{\text{Å}}\,MV)$ | 0.4 |
| $\beta$ | $-Q^2E^2$ | $1.75 \; 10^{11} \; eV/(u\,MV^2)$ | 0.1 |
| $\theta$ | $-QE^3$ | $6.24 \; 10^{14} \; \sqrt{\text{Å}}\,eV/(MV^3\sqrt{u})$ | << 0.1 |
| $\xi$ | $-E^4$ | $3.26 \; 10^{13} \; cm\,eV/MV^4$ | << 0.1 |
| $\phi$ | $-Q^4$ | $1.2 \; 10^{-3} \; eV/(u^2\text{Å})$ | << 0.1 |

Table S3.2: List of the coefficients contained in Eqn. (S3.2) as computed from DFT total energies. Furthermore we show the relative energy contribution from each term for an applied electric field of 8 MV/cm and corresponding Q phonon amplitude.

**S4 – Simulation of the nonlinear optical properties**

The optical properties of SiC were simulated solving one dimensional Maxwell's equations with the finite difference time domain method (1D-FDTD). The discretization of time and space was done according to the Yee Grid, which is suitable for the solution of Maxwell's equation in the absence of free charges, since it intrinsically satisfies the divergence equations $\vec{\nabla} \cdot \vec{B} = 0$ and $\vec{\nabla} \cdot \vec{E} = 0$. The curl equations were explicitly implemented in the FDTD loop, and perfectly absorbing boundary conditions at the end of the grid were used. Being the sample non-magnetic, the constitutive equation for the magnetic field was $\vec{B} = \mu_0 \vec{H}$.

The equilibrium optical properties of SiC were introduced through the constitutive equation $D = \epsilon_0 E + P$. The polarization in the sample was calculated from $P = \epsilon_0 (\varepsilon_\infty - 1)E + \Omega_{TO}\sqrt{\epsilon_0 (\varepsilon_0 - \varepsilon_\infty)}Q$, where $\epsilon_0$ denotes the vacuum permittivity. This equation is equivalent to that in the main text, where $Z^*$ has been expressed in terms of experimentally measurable quantities. The values for the phonon eigenfrequency $\Omega_{TO}$, the damping coefficient $\Gamma$, and the static dielectric function $\varepsilon_0$ were extracted from a Lorentz fit to the static reflectivity data (Ref 12 in the main text), using as a fixed parameter the value of $\varepsilon_\infty$ found in the literature[3]. Similarly, the dynamics of the phonon coordinate was introduced in the FDTD loop with the equation of motion $\ddot{Q} + \Gamma\dot{Q} + \Omega_{TO}^2 Q = \Omega_{TO}\sqrt{\epsilon_0 (\varepsilon_0 - \varepsilon_\infty)}E$.

The non-equilibrium optical properties were calculated using the Hamiltonian shown in Supplementary S3. Considering only the $\alpha$ and $\beta$ expansion terms, which are shown to be the leading ones in Supplementary S3, the Hamiltonian can be written as a function of experimentally measurable quantities in the form

$$H = \frac{1}{2}\Omega_{TO}^2 Q^2 - \frac{1}{2}\epsilon_0(\varepsilon_\infty - 1)E^2 - \Omega_{TO}\sqrt{\epsilon_0(\varepsilon_0 - \varepsilon_\infty)}QE - \alpha Q^3 E - \beta Q^2 E^2$$

And the nonlinear polarization can be expressed as

$$P = -\frac{\partial H}{\partial E} = \epsilon_0(\varepsilon_\infty - 1)E + \Omega_{TO}\sqrt{\epsilon_0(\varepsilon_0 - \varepsilon_\infty)}Q + \alpha Q^3 + 2\beta Q^2 E$$

which is the equation implemented in the FDTD loop. The force acting on the oscillator $Q$ derived from the same Hamiltonian is

$$F_q = -\frac{\partial H}{\partial Q} = -\Omega_{TO}^2 Q^2 + \Omega_{TO}\sqrt{\epsilon_0(\varepsilon_0 - \varepsilon_\infty)}E + 3\alpha Q^2 E + 2\beta QE^2$$

and therefore the equation of motion for the phonon coordinate becomes

$$\ddot{Q} + \Gamma\dot{Q} + \Omega_{TO}^2 Q = \Omega_{TO}\sqrt{\epsilon_0(\varepsilon_0 - \varepsilon_\infty)}E + 3\alpha Q^2 E + 2\beta QE^2$$

which was discretized and implemented in our FDTD simulations.

To simulate the non-linear pump-probe response of the sample, sets of three simulations were performed for each pump-probe delay. The first simulation was considering both the pump (strong) and the probe (weak) field impinging on the sample, and the reflected field $E_{PP}$ was being recorded. A second and a third simulation were then performed, considering only the pump and only the probe field, respectively, therefore calculating the reflected fields $E_{Pump}$ and $E_{probe}$.

The field reflected from the sample in the pumped state was then calculated as $E_{On} = E_{PP} - E_{Pump}$, after making sure that the probe field was weak enough not to drive the system in a nonlinear regime. This was done checking that the reflectivity calculated from $E_{probe}$ was identical to that calculated without nonlinearities.

The reflected field from the unperturbed sample was then calculated as $E_{Off} = E_{probe}$.

From these fields it was possible to obtain $\frac{\Delta_r}{r} = \frac{E_{on} - E_{off}}{E_{off}}$ and calculate the reflectivity in the same way as detailed in Supplementary S2 for the experimental data.